\documentclass[conference]{IEEEtran}
\IEEEoverridecommandlockouts
\usepackage{fancyhdr}
\usepackage{cite}
\usepackage{graphicx}
\usepackage{amsmath,amssymb,amsfonts}
\usepackage{algorithm}
\usepackage{algpseudocode}
\usepackage{multirow}
\usepackage{textcomp}
\usepackage{tabularx}
\usepackage{comment}
\usepackage{xcolor}
\usepackage{booktabs}
\usepackage{enumitem}
\usepackage{array}
\usepackage{url}
\usepackage{hyperref}
\usepackage{pifont}
\usepackage{tikz}

\def\BibTeX{{\rm B\kern-.05em{\sc i\kern-.025em b}\kern-.08em
    T\kern-.1667em\lower.7ex\hbox{E}\kern-.125emX}}

\begin{document}

\title{VERA: Reinforcement Learning for Dynamic Memory Scaling of HPC Workloads in Kubernetes}

\author{
\IEEEauthorblockN{Ade Pramono}
\IEEEauthorblockA{
KTH Royal Institute of Technology\\
Sweden\\
adep@kth.se
}
\and
\IEEEauthorblockN{Jie Ren}
\IEEEauthorblockA{
William \& Mary\\
USA\\
jren03@wm.edu
}
\and
\IEEEauthorblockN{Ivy Peng}
\IEEEauthorblockA{
KTH Royal Institute of Technology\\
Sweden\\
ivybopeng@kth.se
}
}

\maketitle

\begin{abstract}
Memory over-provisioning results in resource underutilization when HPC workloads run on Kubernetes. The default Vertical Pod Autoscaler~(VPA) cannot anticipate phase-driven memory spikes for first-run HPC jobs. In this work, we present a reinforcement learning~(RL) recommender called VERA that formulates vertical memory scaling as a Markov Decision Process and trains an agent on 3{,}353 real Prometheus traces. Evaluated on a live Google Kubernetes Engine~(GKE) cluster using LAMMPS, graph analytics, in-memory analytics, and MLPerf 3D-UNet, the RL agent reclaims 31.6\,\% of the available memory headroom and incurs at most one OOM event while VPA reclaims $-7.9\,\%$ over the same runs, raising memory provisioning, and its recommendation would have been insufficient to avoid OOM in 30 runs. The results demonstrate that an observation-driven RL recommender could outperform retrospective heuristics for dynamic memory scaling.
\end{abstract}


\section{Introduction}
\label{sec:intro}

Memory utilization is a critical factor in the performance and cost-efficiency of HPC workloads. Under-provisioning memory leads to OOM events that terminate jobs prematurely, while over-provisioning wastes resources and increases cost, and motivate for emerging memory systems, such as hybrid memory and disaggregated memory systems in recent years~\cite{peng2016exploring,peng2018characterizing,peng2020demystifying,wahlgren2023quantitative}. Kubernetes is increasingly the common control plane along the edge-cloud-HPC continuum, however, memory management primitives designed for cloud-native microservices translate poorly to the characteristics of HPC workloads, which are often tightly coupled and exhibit diverse memory usage patterns~\cite{beltre2019,medeiros2023kub}. The Kubernetes Vertical Pod Autoscaler~(VPA) is designed to adjust container memory limits based on historical usage patterns, but its default policy is ill-suited for the dynamic and diverse memory profiles of HPC workloads. VPA's recommender observes a rolling histogram of memory consumption and recommends the 90th percentile~(P90) plus a 15\,\% static safety margin~\cite{vpa, selftune2023}. While sufficient for stateless web microservices, this design exposes two critical limitations on HPC workloads.

The first limitation is a workload-agnostic policy. A fixed percentile threshold is appropriate for stationary workloads but fails across diverse HPC memory archetypes: workloads with monotonically growing memory usage require a continuously rising limit; workloads with a sawtooth pattern require a limit set to the peak of each cycle. No single static policy handles all archetypes simultaneously. The second limitation is the inability to adapt to quick phase transitions. HPC jobs frequently exhibit initialization bursts (e.g., graph building, data pre-processing) that complete in seconds and are therefore underrepresented or absent in multi-day histograms~\cite{peng2021holistic,li2023analyzing}. VPA's recommender targets historical averages rather than worst-case future states, making it structurally incapable of anticipating phase-driven memory spikes. The consequences are severe in both directions. Under-provisioning triggers OOM terminating all processes and discarding the entire job run. Over-provisioning consumes cloud memory budget that could otherwise serve additional jobs.

In this work, we formalize memory scaling as a Markov Decision Process with a continuous, proportional action space that rescales the memory limit by a learned factor. We extract a 14-dimensional observation vector from Prometheus metrics. We also design a six-component reward structure that penalizes OOM events while rewarding waste reduction and aligning the scaling direction with observed memory trends. Moreover, the reward applies an init-state guard during the initialization phase, when the metrics are not yet informative. We use PPO to train the RL agent called VERA for \textit{VE}rtical \textit{R}einforcement \textit{A}utoscaler. 

We evaluate VERA on a live Google Kubernetes Engine~(GKE) cluster. We collect 4{,}790 execution traces from HPC workloads, including LAMMPS, Graph Analytics kernels (triangle counting, PageRank, and connected components), Spark-based In-Memory Analytics, and the machine learning application MLPerf 3D-UNet, running in a real cloud environment. The traces cover a range of memory usage patterns from representative workloads in scientific, graph, and machine learning domains. We compare the RL agent with the default VPA recommender on three metrics: OOM avoidance, memory waste reduction, and utilization. Across 100 runs per agent version, the RL agent reclaims 31.6\% of the available headroom against VPA's $-7.9\%$, which raises the provisioned limit above its original value on average. The VPA recommendation would have been insufficient to avoid OOM in 30 runs, while only at most one OOM event occurs with the RL agent. These results demonstrate that an observation-driven RL recommender can outperform retrospective heuristics for dynamic memory scaling of HPC workloads.

In summary, our contributions are as follows:
\begin{itemize}
    \item We formalize memory scaling as an MDP and design an RL agent called VERA that learns to scale memory limits autonomously.
    \item We design a six-component reward structure that penalizes OOM events while rewarding waste reduction aligned with the observed memory trends.
    \item We collect and analyze 4{,}790 execution traces from diverse HPC workloads, covering a range of memory usage patterns.
    \item We evaluate the RL agent against the default VPA recommender on a live GKE cluster, demonstrating significant improvements in memory waste reduction 31.6\% against VPA's $-7.9\%$,.
\end{itemize}

\section{Background and Motivation}
\label{sec:bg}

\subsection{Memory Characteristics of HPC Workloads}
The memory consumption of HPC workloads is often \emph{phase-structured}. Autoscaling tools designed for stateless microservices in the cloud face three difficulties when applied to HPC batch jobs. First, HPC workloads are tightly coupled through MPI communication and synchronization, whereas cloud-native workloads are loosely coupled stateless replicas. Second, HPC jobs rely on application-level checkpoint/restart for fault tolerance and otherwise fail as a unit and discard all progress, whereas cloud workloads tolerate stateless restart and degrade gracefully. Third, an OOM event terminates an entire HPC job and loses all ranks, whereas a cloud-native deployment loses a single replica while the others continue serving.

\subsection{Kubernetes Vertical Pod Autoscaler}
Every container in Kubernetes declares two memory fields: a \emph{request} and a \emph{limit}. The \emph{request} is the minimum guaranteed allocation, used by the scheduler to decide which node can host the pod. The \emph{limit} is the maximum allowed allocation, enforced at runtime by the Linux \emph{cgroup} subsystem. When a container's total memory consumption exceeds its limit, the Linux OOM killer terminates the container.


The Kubernetes VPA attempts to automate the selection of request and limit values. It comprises three components. The \emph{Recommender} monitors historical pod resource usage via Prometheus, builds a histogram over an eight-day observation window, and targets the 90th percentile of observed memory with an additional $15\%$ safety margin. The \emph{Updater} periodically checks whether running pods deviate significantly from the current recommendation and, if so, evicts them so that updated values can be applied. The \emph{Admission Controller} intercepts pod creation requests and injects the latest recommended values into the pod spec before scheduling.

The current VPA has two fundamental limitations for HPC workloads. First, its recommendations are derived from historical usage, but single-run HPC jobs have no prior history. VPA can therefore produce a useful recommendation only after a job has already run, making it \emph{retrospective} rather than adaptive to the current execution. Second, applying a new recommendation requires evicting and restarting the pod; for HPC jobs that run for hours without checkpointing, this discards all completed computation. Kubernetes Enhancement Proposal~1287 (KEP-1287)~\cite{kep1287} introduces the ability to update a container's resource limits at runtime without restarting the pod. 


\subsection{Reinforcement Learning for Autoscaling}
Traditional autoscaling mechanisms, including Kubernetes' default VPA and HPA, rely on threshold-based heuristics that react only after resource utilization crosses a predefined limit. These approaches require application-specific expertise and manual threshold tuning, and they adapt poorly to dynamic workload behavior. Resource allocation is, however, fundamentally a sequential decision-making problem in which each action influences future system states, which makes it a natural fit for reinforcement learning~(RL). Consequently, there is growing interest in RL agents that learn scaling policies directly from interaction with the environment, optimizing cumulative long-term objectives that are often needed to combine conflicting goals, such as sustained performance and cost efficiency.

Proximal Policy Optimization~(PPO) is a policy gradient method that optimizes a surrogate objective with a clipped probability ratio to ensure stable updates~\cite{schulman_ppo}. PPO has emerged as the preferred algorithm for autoscaling problems~\cite{gymhpa,santos2025gwydion,zhang2025kiss}. 
The combination of stability, sample efficiency, and implementation simplicity makes PPO well suited to the noisy state spaces in Kubernetes environments. 

\section{Design}
\label{sec:design}
\begin{figure}[bt]
    \centering
    \includegraphics[width=\linewidth]{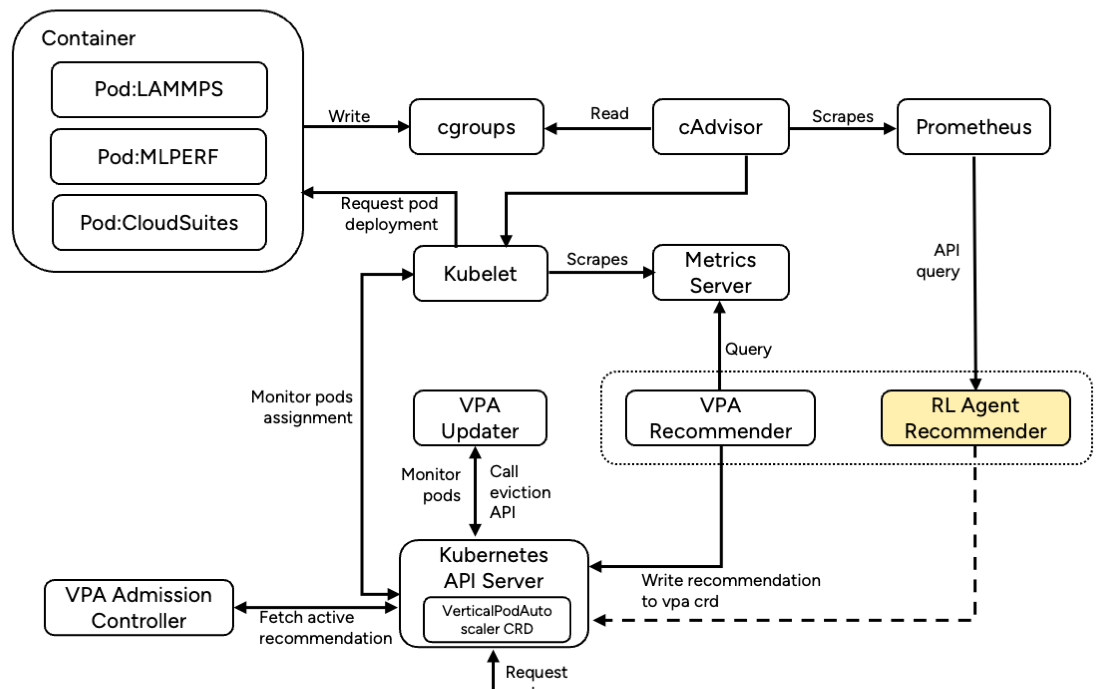}
    \caption{The overall architecture of integrating an RL recommender for vertical memory scaling in Kubernetes's VPA.}
    \label{fig:arch}
\end{figure}
The overall architecture for integrating an RL agent into the Kubernetes autoscaler is illustrated in Fig.~\ref{fig:arch}. The Deployment Controller requests pod creation from the Kubernetes API Server, which holds all cluster state including the VerticalPodAutoscaler CRD. The Kubelet receives the pod assignment and deploys the workload containers, which write their memory consumption directly to cgroups. cAdvisor reads from the cgroups and collects container metrics. The default VPA Recommender queries these metrics through the Metrics Server to produce its memory recommendation, whereas the RL Agent queries Prometheus directly. Because Prometheus scrapes cAdvisor at a finer granularity than the Metrics Server exposes, the agent obtains a richer and more timely observation signal. 
To train and deploy the RL agents, we adopt a sim-to-real pipeline~\cite{santos2025gwydion}. Finally, we propose a composite metric combining waste reduction and OOM avoidance to guide the model selection from a grid of trained RL agents.

\subsection{Reinforcement Learning Environment}
\label{sec:rl}
We model the vertical memory scaling task in Kubernetes pods as a discrete-time Markov Decision Process~(MDP). At each decision step~$t$, an RL agent observes the current state $s_t \in \mathcal{S}$, selects a scaling action $a_t \in \mathcal{A}$, receives a scalar reward $r_t \in \mathbb{R}$, and transitions to the next state $s_{t+1}$. 

The episode advances at 2-second intervals, matching Prometheus scrape resolution. A coarser frequency (e.g., 10\,s) would miss sub-10-second memory spikes while a finer frequency cannot be supported by \textit{cAdvisor}, whose refresh cycle is 10--15\,s. An episode begins at the first Prometheus reading and ends when the pod either completes naturally or is killed by an OOM event. This maps well to HPC jobs and ensures the agent is rewarded across the full memory lifecycle.


\subsubsection{Action Space}
A discrete action set (e.g., $\{-10\%;+10\%\}$) cannot express fine-grained responses. We model action $a_t \in [-1, +1]$ as a continuous adjustment factor applied to the current memory limit $L_t$:
$\hat{L}_{t+1} = \mathrm{clamp}\!\left(
    \hat{L}_t + a_t \cdot \alpha \cdot \hat{L}_t,\;
    L_{\min},\; 2 \cdot L_0
\right)$, where $\alpha = 0.10$ is the maximum adjustment fraction. $L_{\min} = 64$\,MB provides a hard floor below which OOM is certain for most realistic workloads. The ceiling $2 \cdot L_0$ prevents the agent from expanding too much over the operator-set limit, which would waste cluster capacity without benefit.

Proportionality (scaling by $\hat{L}_t$, not $L_0$) provides natural deceleration. When the limit approaches the floor, each step adjusts by a smaller absolute amount, preventing oscillation around the floor. For instance, a pod approaching OOM at 5\,MB/s needs a small positive adjustment, not a maximum expansion. Continuous $a_t \in [-1,+1]$ lets the agent express both direction and magnitude of adjustment simultaneously.

A value of $\alpha = 0.10$ means the agent can move the limit by up to 10\,\% of its current value per 2-second step. Three consecutive maximum-trim steps bring the limit to $0.85^3 \approx 61.4\,\%$ of its
current value, reclaiming meaningful memory within six seconds of a usage drop. Conversely, three consecutive maximum-expand steps can absorb a sudden 52\,\% memory spike ($1.15^3 \approx 1.52$), covering the spike magnitudes observed in many workloads. In contrast, VPA applies its recommendation in a single step on pod restart and cannot make in-episode adjustments.

\subsubsection{State and Observation Space}
In the memory scaling problem, a single measurement of current memory usage does not reveal whether memory need is growing, stable, or declining, which is critical for proactively adjusting the memory limit. To address this partial observability problem, we leverage carefully selected feature engineering in the observation vector. We select features into the observation vector based on three criteria. First, \texttt{Memory relevance}, a feature must directly reflect the pod's memory state or pressure. Features that only loosely correlate with memory behaviour are excluded. Second, \texttt{Markov sufficiency}, a feature must contribute information about the current memory phase that cannot be inferred from the remaining features alone. Third, \texttt{deployment symmetry}, a feature must be observable at inference time from the same Prometheus/cAdvisor pipeline used during training. 

Based on the criteria, we select 14 features. All features are
normalized by the pod's original Kubernetes memory limit $L_0$ (the limit set by the cluster operator), ensuring the policy generalizes across pods of different sizes. We categorize these features into five signal types: OOM-critical signal ($f_0$), OOM precursor signal($f_6$), trend signals($f_7$--$f_9$, $f_{12}$--$f_{13}$), limit signals($f_1$--$f_5$), and CPU context signals($f_10$--$f_11$). 

\begin{itemize}[nosep,leftmargin=*]
    \item $f_0$: \texttt{memory\_working\_set} measures working-set bytes (RSS + active file-backed pages). This is the quantity enforced by the Linux cgroup OOM killer, and thus, the primary safety signal.
    \item $f_1$: \texttt{memory\_rss} measures non-reclaimable pressure (resident set size) and distinguishes from reclaimable size.
    \item $f_2$: \texttt{memory\_usage} is the total memory footprint including page cache.
    \item $f_3$: \texttt{memory\_cache\_ratio} is the fraction of reclaimable page cache over the memory footprint.
    \item $f_4$: \texttt{memory\_limit\_kube} is a cgroup-enforced limit from kube-state-metrics. It represents the hard upper limit which the kernel will OOM-kill the container.
    \item $f_5$: \texttt{memory\_request\_kube} is the minimum memory guaranteed to the pod by the Kubernetes scheduler.
    \item $f_6$:\texttt{memory\_failures\_rate} measures the rate of denied kernel memory allocation requests. This fires before the container working set ($W_t$) breaches $L_t$, giving the agent an early-warning signal to act before OOM is triggered.
    \item $f_7$: \texttt{agent\_limit} is the agent's last recommended limit, providing self-awareness of prior action.
    \item $f_8$: \texttt{usage\_trend} is the rolling slope of working-set $= (W_t - W_{t-1}) / L_{\text{orig}}$. It is initialized to $-1.0$. This feature addresses Markov insufficiency by encoding the direction of memory demand.
    \item $f_9$: \texttt{utilization\_ratio} $= W_t / L_t$ is the primary efficiency signal.
    \item $f_{10}$: \texttt{cpu\_limit\_kube} provides CPU/memory coupling context.
    \item $f_{11}$: \texttt{cpu\_usage\_rate} is the current CPU utilization. It distinguishes CPU-bound from memory-bound phase and acts as a phase transition signal. 
    \item $f_{12}$: \texttt{idle\_steps\_norm} measures the consecutive steps where $W_t < 0.15 L_{\text{orig}}$. It enables trimming during post-computation idle phases. 
    \item $f_{13}$: \texttt{episode\_peak\_norm} is the maximum $W_t$ observed so far in the episode, used for worst-case planning.
\end{itemize}

\noindent In this work, we engineered features over raw time series, instead of feeding into a recurrent policy such as an LSTM~\cite{hausknecht2015drqn}. LSTM training is substantially less stable than MLP training at the 1--5M timestep budgets used in training. We also explored frame stacking~\cite{mnih2015human} as a lightweight alternative for adding temporal context, but its results did not consistently outperform the Markov baseline.

A few engineered features are critical for solving ambiguity. For instance, (\texttt{idle\_steps\_norm}) and (\texttt{episode\_peak\_norm}) were added to solve the ambiguity where memory usage near zero cannot distinguish between ``workload not yet started'' and ``agent successfully trimmed the limit''. Without these features, the reward function cannot differentiate the two states, causing the efficiency signal to misfire during the init phase. Another example is (\texttt{cpu\_usage\_rate}), which provides the primary init-phase signal where a CPU utilization below a threshold $\tau_{\text{cpu}} = 0.20$ of the container's CPU limit indicates the main process is not yet executing. The threshold $\tau_{\text{cpu}} = 0.20$ was selected empirically and is a
configurable hyperparameter. A lower value risks missing the init phase
for slow-starting workloads, while a higher value risks exiting the
init phase prematurely for workloads with a CPU-active setup stage.


\subsubsection{Reward}
\begin{figure}[bt]
    \centering
    \includegraphics[width=1\linewidth]{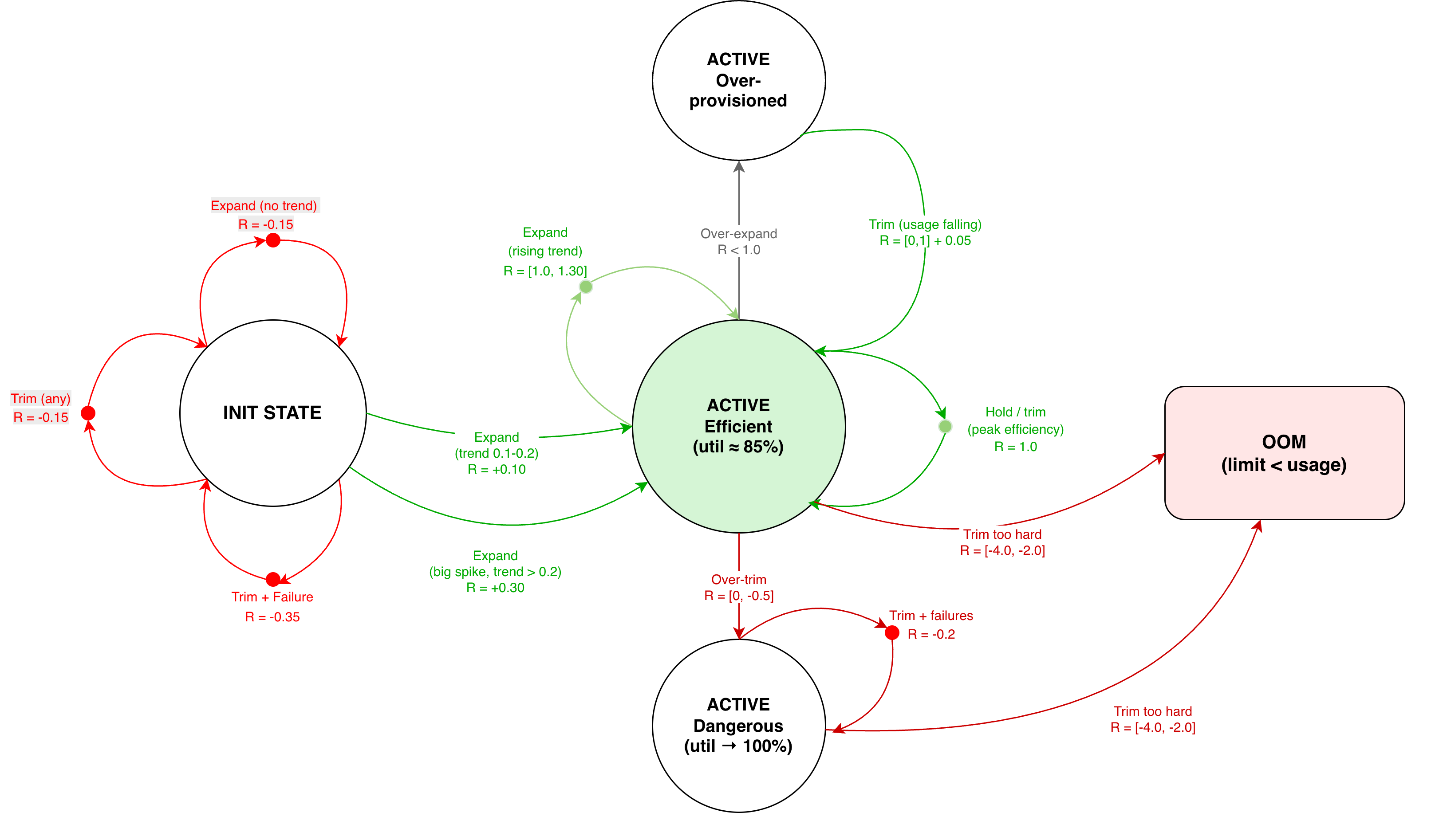}
    \caption{Reward function viewed as a state machine. Green transitions carry positive reward; red transitions carry penalties and represent unsafe or wasteful actions.}\label{fig:reward-fsm}
\end{figure}
We design the reward function to simultaneously encode two objectives that are in direct tension~\cite{sutton2018rl}: \texttt{Safety}, where OOM kills must be heavily penalized as an OOM event crashes the HPC job and wastes all compute time already invested; \texttt{Efficiency}, where the agent must be incentivized to trim the limit toward actual usage to reduce over-provisioned memory waste that could be used to serve other pods.


Fig.~\ref{fig:reward-fsm} illustrates in a finite-state machine that sees the reward function as a whole through the lens of the operational states a pod can occupy. The agent must navigate four states: the \textit{initialization state}, where memory usage is near zero and the computation has not yet started; the \textit{active efficient state}, where utilization is close to the target (85\,\%) and the agent earns its highest reward; the \textit{active over-provisioned state}, where the limit is set higher than demand and memory is being wasted; and the \textit{active dangerous state}, where utilization is approaching 100\,\% and an OOM kill is imminent. The target zone corresponds to a memory limit set 10--15\% above the current working-set: tight enough to recover wasted memory but with sufficient memory headroom to absorb short-term spikes before the OOM killer fires. 

The goal of the reward function is to guide the agent toward the active efficient state and keep it there. From the initialization state, the agent is discouraged from trimming prematurely and rewarded for expanding ahead of a detected memory spike, transitioning cleanly into the efficient state when the workload starts. From the over-provisioned state, small trim bonuses nudge the agent back toward the efficient target. From the dangerous state, the failure pressure penalty and the OOM penalty together create a strong barrier against further trimming. Crossing into the OOM state terminates the episode, making it the absorbing failure state the agent must avoid at all costs. This structure motivates us to design the reward function into six components, each of which governs one or more transitions in the diagram. Table~\ref{tab:reward} summarizes the reward values assigned to each transition. Strong positive rewards reinforce maintaining the efficient zone; large negative penalties escalate as the policy approaches or triggers OOM, calibrated to the asymmetric cost of OOM events~(complete job loss) versus over-provisioning~(wasted budget).



\begin{table}[bt]
  \centering
  \caption{Phase-Aware Reward State-Machine Transitions}
  \label{tab:reward}
  \begin{tabular}{p{4.2cm}c}
    \toprule
    \textbf{Transition} & \textbf{Reward} \\
    \midrule
    Init: Expand with strong rising trend & $[+1.0,\,+1.3]$ \\
    Init: Expand (trend $0.1$--$0.2$, large) & $+0.30$ \\
    Init: Expand (trend $0.1$--$0.2$, small) & $+0.10$ \\
    Init: Expand (no trend detected)       & $-0.15$ \\
    Init: Trim (any) $|Y|$                        & $[-0.15,-0.25,-0.50,-0.75]$  \\
    Init: Trim + allocation failures        & $-0.35$ \\
    \midrule
    Efficient: Hold / trim at peak          & $+1.0$ \\
    Efficient $\to$ Over-provisioned        & $< 1.0$ \\
    Over-provisioned: Trim (falling usage)  & $[0,\,1]+0.05$ \\
    Efficient: Over-trim                    & $[0,\,-0.5]$ \\
    Efficient $\to$ Dangerous: trim+fail    & $-0.2$ \\
    Dangerous / Efficient: Trim too hard    & $[-4.0,\,-2.0]$ \\
    \midrule
    OOM event ($W_t > L_t$)                 & $\ll 0$ (terminal) \\
    \bottomrule
  \end{tabular}
\end{table}

The six components are evaluated in priority order: higher-priority
components short-circuit lower ones. The total reward is clipped to $[-9.0,\;+1.5]$ to avoid unstable policy reward. The first priority component is \texttt{OOM Penalty}. When the agent's limit $\hat{L}_t < W_t$ (working set exceeds the limit):
\begin{equation}
R_{\text{oom}} = -2.0 - 2.0 \cdot \min\!\left(
    \frac{W_t - \hat{L}_t}{W_t},\;1.0
\right) \;\in\; [-4.0,\;-2.0].
\label{eq:oom}
\end{equation}

This component returns immediately, short-circuiting all other components so that no efficiency or trend signal can soften an OOM step. We chose a graduated penalty rather than a fixed value because OOM severity varies across workloads and a constant signal would produce uninformative policy gradients. A slight OOM yields $-2.0$, while a catastrophic OOM where the deficit reaches 100\,\% of usage yields $-4.0$.

The environment terminates the episode when an OOM occurs during the active phase ($\texttt{episode\_peak\_norm} > 0$). Without termination, the agent
could exploit the trajectory ``trim aggressively $\to$ trigger OOM $\to$
recover slowly $\to$ earn sustained efficiency reward'', which is net
profitable under the Bellman equation despite the OOM penalty~\cite{sutton2018rl}. Termination closes this exploit by setting all future rewards to zero after an OOM kill, making aggressive trimming a dead end.

The second highest priority component is \texttt{Cold-Start Guard}. During the init phase ($\texttt{idle\_steps\_norm} > 0 \;\wedge\;
\texttt{cpu\_usage\_rate} < 0.2$):
\begin{equation}
R_{\text{init}} = 0.0 \quad (\text{neutral; no incentive to trim or expand}).
\label{eq:init}
\end{equation}

Without this guard, the efficiency reward in Component~3 would observe near-zero utilization in the initialization state and incorrectly penalize the agent for over-provisioning, incentivizing aggressive trimming before the job has even started its active computation phase. When memory demand then spikes at the transition to the active state, the limit would already be too low, causing an immediate OOM. The neutral signal means holding the limit steady is the correct action during init. 

The third component is \texttt{Efficiency Reward} that applies to active phases.
\begin{equation}
R_{\text{eff}} =
\begin{cases}
1.0 + \dfrac{\rho_t - \rho^*}{\rho^*}  & \text{if } \rho_t \leq \rho^* \\[6pt]
\max\!\left(0,\; 1.0 - 10(\rho_t - \rho^*)\right) & \text{if } \rho_t > \rho^*
\end{cases}
\;\in\; [0,\;1],
\label{eq:efficiency}
\end{equation}
where $\rho_t = \frac{W_t}{\hat{L}_t}$, $\rho^* = \frac{1}{1 + \delta}$, and $\delta = 0.15$ is the target safety margin, giving $\rho^* = 1/1.15 \approx 0.870$. The asymmetric slope reflects the asymmetric cost structure we observe in practice. On the safe side of the target, the agent has 0.87 utilization units of headroom before reaching complete over-provisioning. On the dangerous side, it has only 0.13 units before hitting the OOM boundary. We therefore apply a $10\times$ steeper slope on the right side of $\rho^*$, so that a small drift toward OOM costs the agent far more reward than an equivalent drift toward over-provisioning, naturally biasing the policy toward safety.

We set the target safety margin to $\delta = 0.15$, which aligns with the default \texttt{recommendation-margin-fraction} of 15\,\% used by the Kubernetes VPA~\cite{vpa} and corroborated by industry practice reported in SelfTune~\cite{selftune2023}. This margin is large enough to absorb cAdvisor polling jitter across a 2-second scrape interval and small Spark allocation fluctuations, while remaining tight enough to produce a meaningful waste reduction over VPA's conservative P90 baseline.

The fourth component is \texttt{Trend Alignment Bonus}. We designed this component to reward the agent for moving in the same direction as memory demand. The two directions of memory demand (increase/decrease) carry very different urgency: failing to expand ahead of a rising spike leads directly to OOM, while failing to trim during a usage decline leads to over-provisioning. Both outcomes are undesirable, but OOM is immediately catastrophic whereas over-provisioning is merely wasteful. We therefore apply an asymmetric 6:1 ratio between the expand and trim bonuses, with a maximum of $+0.30$ for expanding into a rising spike and $+0.05$ for trimming during a usage decline, so that the agent learns to prioritize safety over efficiency when the two objectives conflict.
\begin{equation}
R_{\text{trend}} =
\begin{cases}
+0.30 \cdot \min(d_t, 1)  & d_t > 0.2 \;\wedge\; a_t > 0 \\
+0.10 \cdot \min(d_t, 1)  & d_t > 0.1 \;\wedge\; a_t > 0 \\
+0.05 \cdot \min(-d_t, 1) & d_t < -0.1 \;\wedge\; a_t < 0 \\
0 & \text{otherwise},
\end{cases}
\label{eq:trend}
\end{equation}
where $d_t = \texttt{usage\_trend} = (W_t - W_{t-1})/L_0$.

The fifth component is \texttt{Failure Pressure Penalty}. If \texttt{memory\_failures\_norm} $> 0.5$ and $a_t < 0$ (agent is trimming under active OOM pressure), it receives $R_{\text{fail}} = -0.20$. During our experiments, we observed that the OOM penalty in Component~1 alone cannot always prevent the agent from trimming a pod that is already under memory pressure. A page-major-fault rate above a configurable threshold $\tau_{\text{fail}} = 0.5$ indicates that the pod is actively struggling to meet its memory demand, and trimming further in this state risks triggering an OOM kill. We therefore introduce a second-level safety interlock that applies a penalty if the agent trims when \texttt{memory\_failures\_norm} $> \tau_{\text{fail}}$. This component acts as a last line of defense that activates only when Component~1 has not yet fired, and $\tau_{\text{fail}}$ is a configurable hyperparameter. 

The sixth component is \texttt{Init-Trim Penalty}. During the init phase, if $a_t < 0$ (i.e., agent trims), $R_{\text{itrim}} = Y < 0 $. This penalty discourages the agent from reclaiming memory when it is in init state, where the magnitude $Y$ is tuned empirically via a sensitivity study in this work. If the init-phase trimming is too costly, the agent enters the active state with the limit still close to $L_0$, at which point the weak trim signal of at most $+0.05$ from Component~4 is insufficient to bring the limit down to an efficient level before the episode ends. We find $Y = -0.15$ to strike the best balance.

\subsection{Reinforcement Learning Agent}
\label{sec:rlagent}
We adopt an actor-critic network architecture, where the actor (the policy network) proposes memory adjustment actions, and the critic (the value network $V_{\pi}$) estimates expected returns under the actor’s policy. 

\subsubsection{Asymmetric Actor-Critic Architecture}
We keep actor and critic networks separate to ensure that value function updates do not interfere with policy updates through a shared backbone, a known source of instability in actor-critic training~\cite{andrychowicz2020matters}. The policy network uses a small ($(32, 32)$) two-layer MLP to encourage generalization across diverse workload traces. Since the action space is a single scalar that requires minimal parametric capacity, a smaller network is less prone to overfitting across the diverse workload traces in the training set. The critic, on the other hand, needs to memorize a complex, phase-dependent value landscape to produce accurate long-horizon GAE estimates~\cite{schulman2015gae}. Therefore, the value network uses a larger ($(256, 256)$) two-layer MLP to accurately estimate long-horizon returns. 

We choose an MLP policy because the \texttt{idle\_steps\_norm} and \texttt{usage\_trend} features already encode the most critical temporal information in the state vector, removing the need for a recurrent architecture to infer temporal context from raw sequences. We use Tanh activation because they produce outputs bounded in $[-1, +1]$ that aligns naturally with the action space range and avoids the saturation instabilities that ReLU activation can introduce in continuous control settings.  

The observation captures the current memory state well but tells the agent little about how that state has been evolving over recent steps. We address this by frame stacking~\cite{mnih2015human} that concatenates the last $N$ observations into a flat vector of dimension $14 \times (N+1)$, giving the policy a short memory of
recent states. 

We train the RL agent using Proximal Policy Optimization
(PPO)~\cite{schulman_ppo}. The clipped surrogate objective
($\varepsilon = 0.2$) bounds each policy update. Since the OOM reward signal fires infrequently during training, a single catastrophic policy update could destroy accumulated policy quality, and thus training stability from PPO is critical. Also, PPO with a Gaussian policy handles $a_t \in [-1,+1]$ natively without discretization artifacts. Finally, PPO applies multiple gradient epochs to the same batch under its clipped surrogate objective, which improves learning efficiency without requiring additional data collection. This sample efficiency is important when real pod executions take minutes to hours. 
Table~\ref{tab:ppo} lists the PPO hyperparameter configuration. The long rollout buffer ($N=2048$ steps per environment) captures full-episode temporal dynamics. The discount factor $\gamma=0.99$ is chosen because HPC jobs run for thousands of steps, and a low discount value would excessively discount late-episode efficiency gains that represent the majority of a job's wall time.

\begin{table}[bt]
\centering
\small
\caption{PPO hyperparameter configuration.}
\label{tab:ppo}
\resizebox{\columnwidth}{!}{%
\begin{tabular}{lll}
\toprule
Parameter & Value & Rationale \\
\midrule
Learning rate $\eta$     & $3 \times 10^{-4}$  & Adam default; stable for continuous control \\
Rollout steps $N$        & 2048                & Captures full episode temporal dynamics \\
Mini-batch size          & 1024                & 8 mini-batches per update\\
Gradient epochs $K$      & 10                  & Improves sample efficiency; clip prevents collapse \\
Discount $\gamma$        & 0.99                & Long episodes require high discount \\
GAE $\lambda$            & 0.95                & Standard bias-variance trade-off \\
Clip range $\epsilon$    & 0.2                 & Standard PPO; prevents destructive updates \\
Entropy coefficient      & 0.01                & Small exploration bonus \\
Value coefficient        & 0.5                 & Standard value loss weight \\
Max gradient norm        & 0.5                 & Gradient clipping for stability \\
Training timesteps       & $1-3 \times 10^6$ & Per workload; tuned by convergence \\
\bottomrule
\end{tabular}
}
\end{table}


The agent is trained in a sim-to-real pipeline and deployed to the live GKE cluster for evaluation. We collect the dataset by sweeping workload-specific problem-size parameters across memory provisioning tiers.
This stage produces 1,214,600~rows of Prometheus time-series samples and 4,790~labeled execution traces. The dataset is shuffled across workload types and split 70/30 into training and evaluation. 

\subsubsection{Training Hyperparameter Grid and Model Selection}
\label{sec:modelselect}
A total of 96 PPO agents are trained by varying frame-stack depth and training budget. 
All configurations use the same PPO hyperparameters: $\gamma = 0.99$, $\lambda_{\text{GAE}} = 0.95$, clip $\varepsilon = 0.2$, $K = 10$~gradient epochs per rollout, $\eta = 3\!\times\!10^{-4}$~(Adam). The critic network is fixed at
$(256, 256)$ hidden units in all configurations.

We observed that relying on a single scalar metric to select the production agent model is insufficient. A fully trained model with the highest utilization ratio may operate dangerously close to the OOM boundary, while a model with the lowest OOM rate may over-provision excessively. Therefore, we evaluate each trained model using three metrics jointly, each capturing a dimension of the safety-efficiency trade-off: $m_1$: mean utilization ratio $= \overline{W_t / \hat{L}_t}$; $m_2$: Total OOM events per experiment; $m_3$: mean waste ratio $= \overline{(\hat{L}_t - W_t)/L_0}$.

To produce a single comparable score, we apply two complementary normalization schemes. Min-Max normalization rescales each metric relative to the best and worst values observed across all candidates; it assumes a uniform distribution of scores and is sensitive to outlier models. Z-score normalization instead centers each metric on its mean and scales by its standard deviation, which is more robust to outliers but sensitive to the population standard deviation. The final production model is the model on the Pareto front of $(m_2, m_3)$ with the shortest distance to the utopia point under both schemes, so that the selection is robust to either distributional assumption. 

\section{Experimental Setup}
\label{sec:setup}
We evaluate the RL agent on a live GKE cluster. Data collection is performed on GKE Cluster~v1.35.1-gke.1396002 in region \texttt{europe-north1-a}. Training traces are collected on a single \texttt{n2-highmem-8} node (8~vCPUs, 64\,GB RAM), and evaluation traces on an \texttt{n2-standard-16} node (16~vCPUs, 64\,GB RAM). 
At each 5-second interval, the RL agent queries Prometheus, computes action $a_t$ to obtain a new memory limit. 
Meanwhile, the VPA baseline runs with \texttt{updateMode} off, and its emitted recommendations are logged for comparison. Both recommenders run in shadow mode: VERA's computed limit $\hat{L_t}$ is logged rather than applied, so every pod executes under its original limit $L_0$ for the full run, a run is counted as an OOM for a recommender if $W_t$ exceeded that recommender's $L_{rec}$ at any point. The OOM counts are therefore first-crossing counts on a fixed trace comparable between the two recommenders, since both are evaluated against the identical execution. Enforcing the limit requires in-place limit updates via KEP-1287, and in particular whether the kubelet honours limit decreases on a live container, which is left to future work.


We evaluate four agent versions trained with different init-trim penalties, $Y \in \{-0.15, -0.25, -0.50, -0.75\}$. Each version is evaluated over 100 workload runs spanning all four workloads: 50 runs for Graph Analytics (including BFS, PR, and CC applications), 15 for In-Memory Analytics, 18 for LAMMPS, and 17 for MLPerf 3D-UNet. These counts reflect the number of algorithm variants, input configurations, and memory tiers available for each workload.

We use three metrics to evaluate the performance of memory scaling under RL agents and VPA: Memory waste reduction $\Delta W =
\frac{L - \bar{r}}{L - \bar{u}} \times 100\,\%$, Utilization ratio \textit{Util} $=\frac{\overline{W_t}}{L_{\text{rec}}}$, and the number of OOM events when $\overline{W_t}>\overline{L_\text{rec}}$
\noindent where $L$ is the original memory limit, $\bar{r}$ is the recommended memory limit, $\bar{u}$ is the mean memory usage, $\overline{W_t}$ is the mean working set across the episode, and $L_{\text{rec}}$ is the recommended limit. \textit{Util} is particularly relevant because it reflects how tightly the recommended limit tracks the actual demand.
\section{GKE Production Evaluation}
\label{sec:eval}
In this section, we evaluate our RL agents against the default Kubernetes VPA recommender on real workloads, and analyze the agent's behavior on each workload.


\subsection{Overall Performance}
\begin{table}[bt]
\centering
\caption{Three evaluation metrics measured in all workloads. $-$ means the recommendation is unusable.} 
\label{tab:aggregate}
\resizebox{\linewidth}{!}{
\setlength{\tabcolsep}{5pt}
\begin{tabular}{llrrrrrr}
\toprule
\multirow{2}{*}{\textbf{Workload}} &
\multirow{2}{*}{\textbf{RL Model}} &
\multicolumn{3}{c}{\textbf{RL Agent}} &
\multicolumn{3}{c}{\textbf{VPA Baseline (mean)}} \\
\cmidrule(lr){3-5}\cmidrule(lr){6-8}
& & $\Delta W$ & \textit{util} & \textit{\#OOM}
  & $\Delta W$ & \textit{util} & \textit{\#OOM} \\
\midrule
\multirow{4}{*}{CS Graph ($n{=}50$)}
  & 1 & 30.0\% & 0.136 & 0 & \multirow{4}{*}{2.2\%}  & \multirow{4}{*}{0.017} & \multirow{4}{*}{7} \\
  & 2 & 28.6\% & 0.142 & 0 & & & \\
  & 3 & 17.5\% & 0.129 & 0 & & & \\
  & 4 & -0.3\% & 0.113 & 0 & & & \\
\midrule
\multirow{4}{*}{CS In-Mem ($n{=}15$)}
  & 1 & 62.7\% & 0.462 & 0 & \multirow{4}{*}{-}  & \multirow{4}{*}{-} & \multirow{4}{*}{13} \\
  & 2 & 67.2\% & 0.514 & 0 & & & \\
  & 3 & 61.3\% & 0.474 & 0 & & & \\
  & 4 & 54.1\% & 0.467 & 0 & & & \\
\midrule
\multirow{4}{*}{LAMMPS ($n{=}18$)}
  & 1 & 25.5\% & 0.404 & \textbf{1} & \multirow{4}{*}{$-51.5\%$} & \multirow{4}{*}{0.298} & \multirow{4}{*}{4} \\
  & 2 & 17.9\% & 0.435 & 0           & & & \\
  & 3 &  3.2\% & 0.419 & 0           & & & \\
  & 4 &  4.6\% & 0.419 & 0           & & & \\
\midrule
\multirow{4}{*}{MLPerf ($n{=}17$)}
  & 1 & 15.4\% & 0.225 & 0          & \multirow{4}{*}{1.5\%}  & \multirow{4}{*}{0.014} & \multirow{4}{*}{6} \\
  & 2 & 16.2\% & 0.245 & \textbf{1} & & & \\
  & 3 &  4.8\% & 0.232 & 0          & & & \\
  & 4 &  4.5\% & 0.277 & 0          & & & \\
\midrule
\multirow{4}{*}{\textbf{Overall ($n{=}100$)}}
  & \textbf{1} & \textbf{31.6\%} & 0.248 & \textbf{1} & \multirow{4}{*}{$-7.9\%$} & \multirow{4}{*}{0.065} & \multirow{4}{*}{30} \\
  & 2          &          30.4\% & 0.268 & \textbf{1} & & & \\
  & 3          &          19.3\% & 0.250 & \textbf{0} & & & \\
  & 4          &           9.6\% & 0.249 & \textbf{0} & & & \\
\bottomrule
\end{tabular}
}
\end{table}

Across all three evaluation metrics, the RL agent consistently outperforms VPA. Table~\ref{tab:aggregate} reports the mean value of each metric per agent version and workload. RL Model~1 ($Y = -0.15$) is the best-performing configuration. On OOM count, the RL agents incur at most one event out of 100~runs, whereas the VPA recommendation would have been insufficient in 30~runs. On waste reduction, VPA records $-7.9\,\%$ overall, meaning that it raises the provisioned memory above the original limit on average rather than reclaiming headroom. The RL agent, in contrast, achieves an overall waste reduction of $31.6\,\%$. As the init-trim penalty $|Y|$ increases, the waste reduction achieved by the RL agent falls from $31.6\,\%$ to $9.6\,\%$ while the OOM count falls to zero, exposing a consistent trade-off between init-phase conservatism and efficiency.

\subsection{Graph Analytic Applications}
\label{sec:graph-results}
\begin{figure}[bt]
    \centering
    \includegraphics[width=1\linewidth,height=0.38\linewidth]{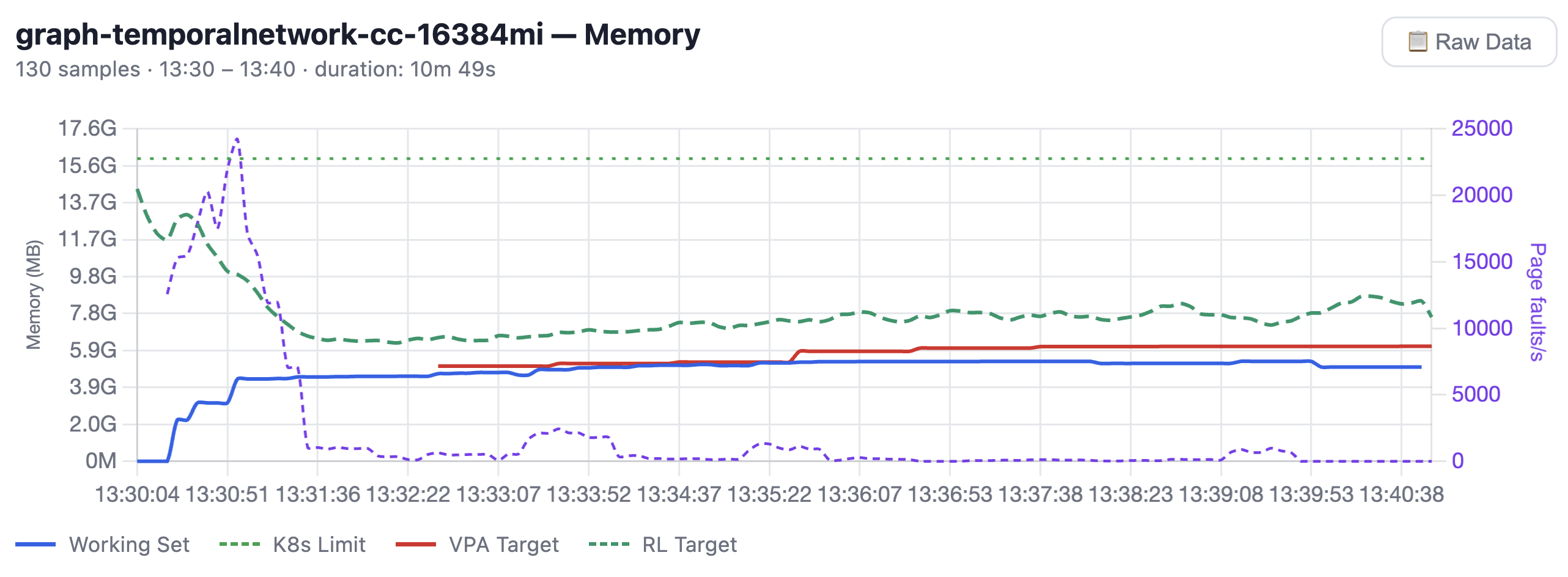}
    \caption{A real trace of a Graph Analytics Connected Components application (16\,GiB tier, Model~1).}
    \label{fig:trace-graph}
\end{figure}

A Graph Analytics job starts with a fast initialization ramp, after which memory grows steadily as algorithms traverse the graph and accumulate intermediate results. Fig.~\ref{fig:trace-graph} presents a real trace. VPA cannot anticipate this growth because it derives its recommendation from historical usage and sets the memory limit to previously observed peaks. As traversal depth and working-set size increase throughout the execution, the actual demand often exceeds this historical ceiling.

We evaluate the graph workload on real-world networks from the Stanford Large Network Dataset Collection (SNAP)~\cite{snapnets}, using the
\emph{temporalnetwork} (\texttt{sx-stackoverflow}) and \emph{web}
(\texttt{web-Google}) graphs. Each algorithm (CC, PR, TC) is deployed across three memory tiers, 8, 16, and 32\,GiB, with the pod's memory request fixed to its limit. Across the 50 runs, the RL agent sets a tighter memory limit while maintaining a better safety record than VPA: the RL agents record no OOM events, whereas the VPA recommendation would have been insufficient in seven runs on average. Waste reduction under the RL agent is 30.0\,\% against VPA's 2.2\,\%.

The most severe VPA failures occur on the \emph{temporalnetwork} dataset family. At the 8\,GiB tier, CC, PR, and TC each record a VPA OOM; at the 16\,GiB and 32\,GiB tiers, PR and TC do so as well. 
The RL agent records zero OOM events across all of these executions. It avoids OOM by adjusting the limit upward continuously in response to the observed growth signal, rather than relying on a fixed historical estimate.


\subsection{In-Memory Analytics}
\label{sec:inmem-results}
\begin{figure}[bt]
    \centering
    \includegraphics[width=\linewidth]{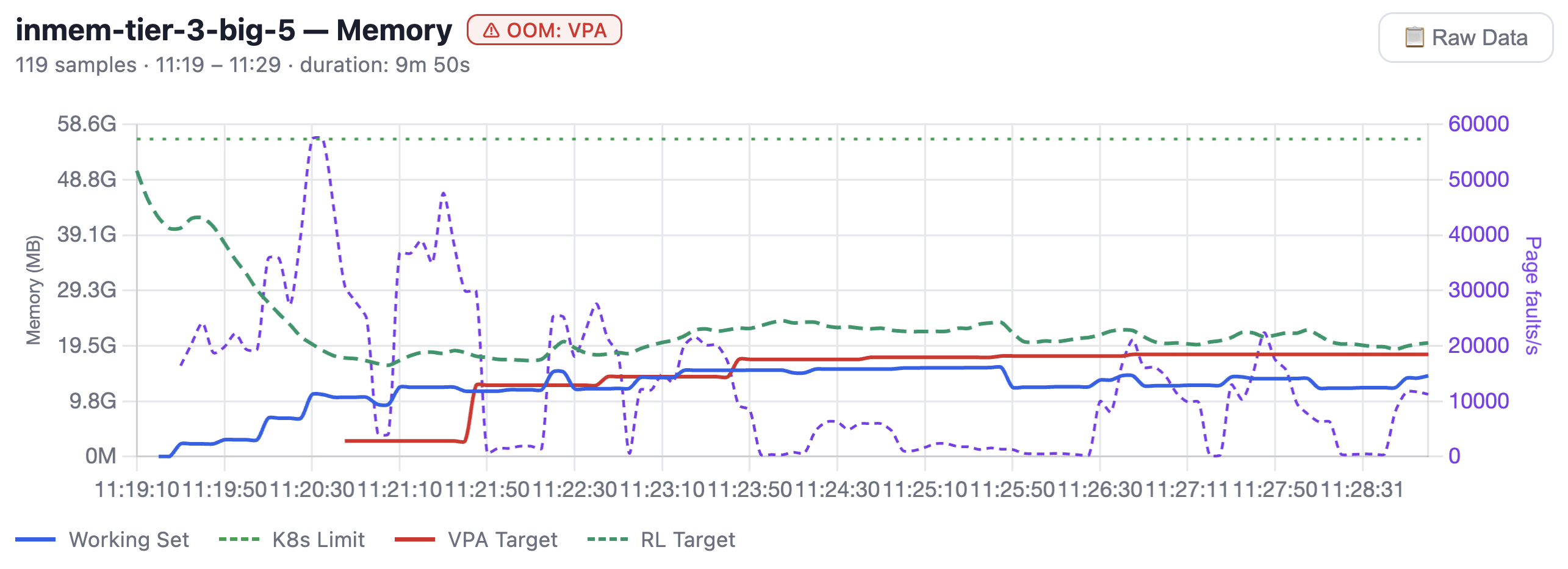}
    \caption{A real trace of In-Memory Analytics (\emph{tier-3-big-5}, Model~1).}
    \label{fig:trace-inmem}
\end{figure}
The In-Memory Analytics workload exhibits the sharpest contrast between RL and VPA. This workload 
has memory usage with no regular period or predictable ceiling. Across 15 runs, VPA records an average of 13 OOM events. The RL agent records zero OOM events in all four versions. Moreover, VPA's cannot reduce any memory waste because it produced no usable recommendation for any in-memory configuration when a workload's burst magnitude is uncapped and irregular. The RL agent achieves waste reductions of 62.7\,\% (Model~1), 67.2\,\% (Model~2), 61.3\,\% (Model~3), and 54.1\,\% (Model~4), with an average memory utilization in 0.462--0.514. 
 
We show a representative trace in Fig.~\ref{fig:trace-inmem}. The working set oscillates throughout the whole run. The RL target begins conservatively near 49~GiB and contracts steadily as demand becomes observable. It settles between 17-24 GiB for the remainder of the run and consistently tracks tightly above each burst peak without causing any OOM event. In contrast, VPA issues its first recommendation at only 2.8\,GiB, a value already 3.8$\times$ below the observed working-set of 10.7\,GiB. By setting the limit around 20\,GiB against an original memory limit of 57\,GiB, the RL agent recovers 37\,GiB of memory resources, achieving a waste reduction of $65\,\%$.

\subsection{LAMMPS}
\label{sec:lammps-results}
\begin{figure}[bt]
    \centering
    \includegraphics[width=\linewidth,height=0.4\linewidth]{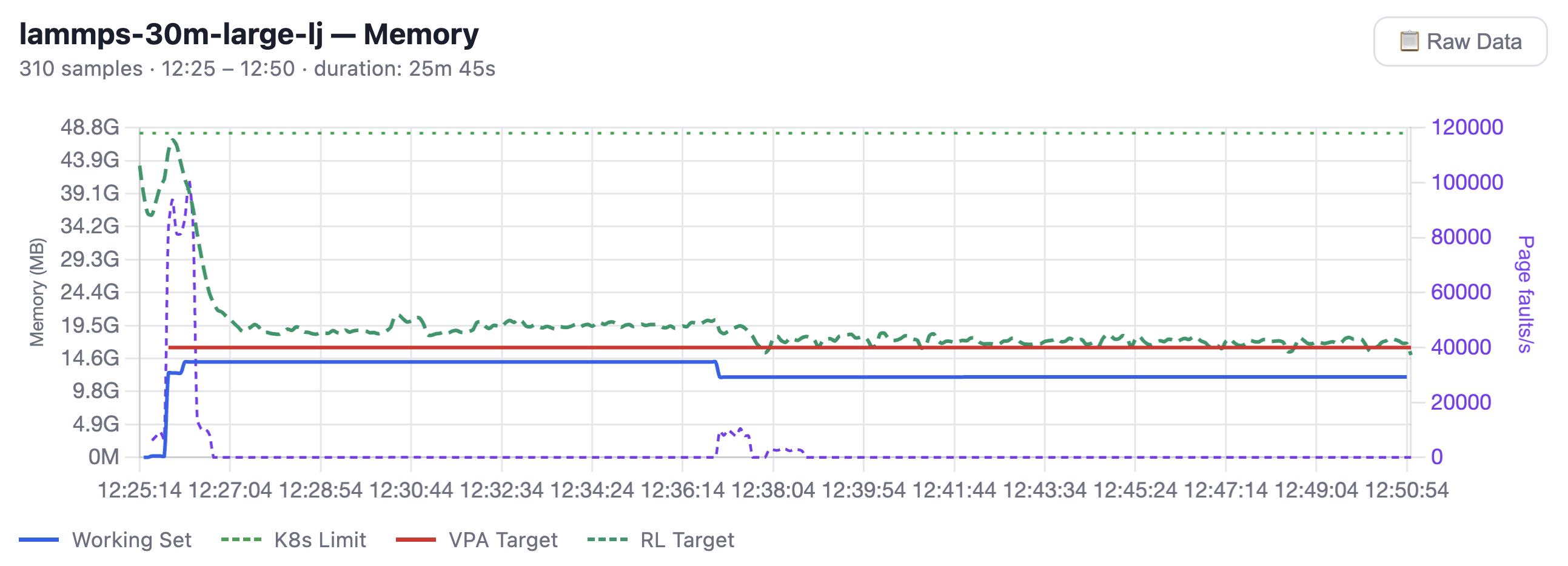}
    \caption{A real trace of LAMMPS MD simulation (\emph{30m-large-lj}, Model~1).}
    \label{fig:trace-lammps}
\end{figure}
LAMMPS runs molecular dynamics simulations whose memory consumption scales with atom count, producing a monotonically rising demand trajectory. The most notable finding on this workload is VPA's negative waste reduction of $-51.5\,\%$. A negative $\Delta W$ indicates that VPA's recommended limit exceeds the originally provisioned limit, aggravating over-provisioning. For a monotonically increasing workload, the running peak always sits at the top of the memory histogram, so VPA continually recommends above the previous provisioning level. The RL agent achieves positive waste reduction in all four models: 25.5\,\%(Model~1), 17.9\,\% (Model~2), 3.2\,\% (Model~3), and 4.6\,\% (Model~4). Utilization under the RL agent rises to 0.404--0.435, against 0.298 for VPA. Fig.~\ref{fig:trace-lammps} shows a representative LAMMPS episode trace in which the RL agent closely tracks the linear growth curve while VPA's recommendation remains constant throughout the run.

On large Lennard-Jones simulations, Model~1 achieves the highest waste reduction of 81.1\,\% for \emph{30m-large-lj}, 75.4\,\% for \emph{40m-large-lj}, and 69.4\,\% for \emph{30m-medium-lj}. In these configurations the agent tracks the predictable linear growth and trims the limit closely above demand. The \emph{real} and \emph{metal} unit-system variants at smaller atom counts yield lower waste reduction, consistent with their smaller memory footprint, which leaves less headroom to reclaim. 
 
On small simulations (\emph{40m-small-lj}), one OOM event is recorded under Model~1 at a 16\,GiB limit. This configuration runs a simulation with an unusually steep early-phase memory growth rate, and the init-trim penalty $Y = -0.15$ is insufficient to hold the limit open during the init window on this trajectory. Models~2, 3, and~4, with stronger penalties, record zero OOM on the same sub-configuration, which confirms that the init-trim penalty is a critical design element for HPC workloads such as LAMMPS. 

\subsection{MLPerf 3D-UNet}
\label{sec:mlperf-results}
\begin{figure}[bt]
    \centering
    \includegraphics[width=\linewidth,height=0.4\linewidth]{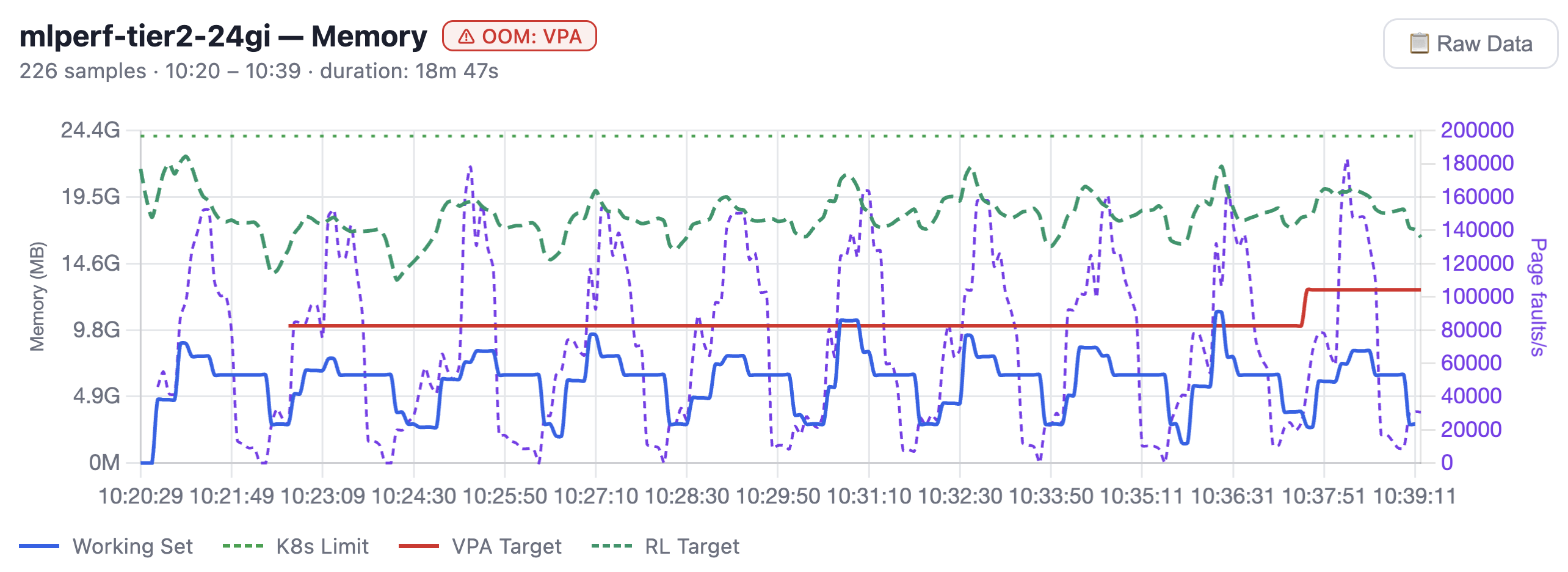}
    \caption{A real trace of MLPerf 3D-UNet workload (\emph{tier2}, 24\,GiB, Model~1).}
    \label{fig:trace-mlperf}
\end{figure}
 
MLPerf 3D-UNet exhibits a repeating sawtooth memory pattern driven by mini-batch loading and framework-level tensor caching. One representative trace is shown in Fig.~\ref{fig:trace-mlperf}. Every burst peak is an OOM risk if the limit falls below it. Across the 17 MLPerf evaluation runs, the VPA recommendation would have been insufficient in six runs, whereas the RL agent records zero OOM events in Models~1, 3, and~4, and one in Model~2. The \emph{tier1}, \emph{tier2}, and \emph{tier3} labels denote 3D-UNet configurations with activation sizes of 1024, 4096, and 7680, respectively. Within \emph{tier2}, VPA records OOM at the 12, 24, and 49\,GiB memory limits while the RL agent records none; the same pattern holds within \emph{tier3} at 32 and 49\,GiB.
 

The RL agents achieve a modest waste reduction on this workload: 15.4\,\% (Model~1), 16.2\,\% (Model~2), 4.8\,\% (Model~3), and 4.5\,\% (Model~4), against VPA's 1.5\,\%. The sawtooth peaks set a hard demand ceiling that the agent cannot predict in advance, so avoiding OOM requires holding a conservative buffer above every peak at all times, which bounds the headroom that can be reclaimed. As shown in Fig.~\ref{fig:trace-mlperf}, the RL agent maintains a buffer above each sawtooth peak, while VPA's recommendation falls below peak demand and would have caused an OOM.

\section{Related Work}
\label{sec:relatedwork}
Rossi et al.~\cite{rossi2019} present RL-based scaling controllers for containerized applications, using DQN to
jointly manage horizontal and vertical scaling with an objective balancing response time, cost, and adaptation frequency. Their work targets CPU and latency for web services rather than memory waste and OOM prevention. 

Gym-hpa~\cite{gymhpa} implements PPO, A2C, and RPPO agents for horizontal autoscaling of microservices in Kubernetes, with the objective of reducing pod counts relative to the default HPA while maintaining acceptable service latency. Gwydion~\cite{santos2025gwydion} extends gym-hpa with a sim-to-real transfer interface and integrated statistical forecasting, showing that simulation-trained agents reach competitive performance at approximately one-tenth of the training cost of online-trained agents. 

KIS-S~\cite{zhang2025kiss} applies PPO to GPU-aware horizontal autoscaling for inference on Kubernetes, reducing P95 latency by up to $6.7\times$ relative to CPU-only baselines and showing that PPO-based scaling policies generalize well across inference workload configurations. Different from these works on optimizing horizontal scaling, this work targets vertical scaling.

AWARE~\cite{qiu2023aware} explored RL in workload autoscaling in production cloud platforms. They proposed an extensible framework for deploying RL agents in production systems and can adapt a learned auto-scaling policy much faster than the existing transfer-learning-based approach for new workloads and significantly reduce SLO violations. 

In addition to RL-based methods, threshold-based reactive scaling approaches are also used to scales up or down a Kubernetes cluster when usage exceeds or falls below a threshold. ARC-V~\cite{medeiros2025arc} implements a three-state heuristic controller (growing, stable, dynamic) with in-place memory limit adjustment. They evaluated on nine HPC applications, and confirms that workload-aware in-place resizing can outperform the default VPA heuristics for HPC workloads. However, their thresholds need to be hand-tuned per workload class and do not adapt from experience, so each new workload type requires manual re-engineering.

\section{Conclusion}
\label{sec:concl}
In this work, we presented an RL-based recommendation framework called VERA for dynamic vertical memory autoscaling on Kubernetes. We formulated the task as an MDP, trained PPO agents offline on 3{,}353 real traces collected from GKE clusters, and evaluated them against the default VPA recommender on a live GKE cluster in LAMMPS, MLPerf 3D-UNet, In-Memory Analytics, and Graph Analytics workloads. The results show that the RL agent prevents OOM more reliably than VPA while recovering substantial memory waste. It incurs at most one OOM event, whereas the VPA recommendation would have been insufficient in 30 runs, and it reduces memory waste by 31.6\,\% while VPA raises the provisioned memory limit by 7.9\,\% on average. Extending coverage to more HPC workloads with diverse phase patterns is the natural next step toward an RL-driven Recommender that complements the default VPA. 

\section*{Acknowledgment}
This research is supported by the European Commission under the Horizon project OpenCUBE (101092984).

\bibliographystyle{IEEEtran}
\bibliography{refs}

\end{document}